\documentclass[reprint,superscriptaddress,amsmath,amssymb,aps,prd,notitlepage,floatfix,nofootinbib,longbibliography,onecolumn]{revtex4-1}

\usepackage{changes}
\usepackage{tensor}     
\usepackage{graphicx}   
\usepackage[
colorlinks=true,        
citecolor=blue,         
linkcolor=blue,         
urlcolor=blue           
]{hyperref}             
\usepackage{bm}         
\usepackage{xcolor}     
\usepackage{lipsum}
\usepackage{color}      
\usepackage[utf8]{inputenc} 
\usepackage[section]{placeins} 
\usepackage{multirow}
\usepackage{orcidlink}
\newcommand{\nc}{\newcommand*} 

\nc{\al}{\alpha}
\nc{\s}{\sigma}
\nc{\kp}{\kappa}
\nc{\dt}{\delta}
\nc{\Dt}{\Delta}
\nc{\Ld}{\Lambda}
\nc{\p}{\partial}
\nc{\Gm}{\Gamma}
\nc{\om}{\omega}
\nc{\Om}{\Omega}
\nc{\rd}{\mathrm{d}}
\def\({\left(}
\def\){\right)}
\def\[{\left[}
\def\]{\right]}
\def\e{\begin{equation}}
\def\q{\end{equation}}
\def\m{\begin{eqnarray}}
\def\n{\end{eqnarray}}
\nc{\Eq}[1]{Eq.~\eqref{#1}}     
\nc{\Eqs}[1]{\eqref{#1}} 
\nc{\Fig}[1]{Fig.~\ref{#1}}     
\nc{\Table}[1]{Table~\ref{#1}}  
\nc{\Sec}[1]{Sec.~\ref{#1}}     
\nc{\Msun}{M_\odot}             
\nc{\fpbh}{f_{\mathrm{pbh}}}    
\nc{\fpbhn}{f_{\mathrm{pbh0}}}    
\nc{\mR}{\mathcal{R}} 
\nc{\seq}{\sigma_{\mathrm{eq}}}
\nc{\ogw}{\Omega_{\mathrm{GW}}}
\nc{\gpcyr}{\mathrm{Gpc}^{-3}\,\mathrm{yr}^{-1}}
\nc{\lvc}{LIGO/Virgo} 
\nc{\SNR}{\mathrm{SNR}} 
\nc{\mmin}{{m_{\mathrm{min}}}}
\nc{\mmax}{{m_{\mathrm{max}}}}
\nc{\Mmin}{{M_{\mathrm{min}}}}
\nc{\fmin}{{f_{\mathrm{min}}}}
\nc{\VT}{\mathrm{VT}}
\nc{\rhoGW}{\rho_{\mathrm{GW}}}
\nc{\vth}{\vec{\theta}}
\nc{\vd}{\vec{d}}
\nc{\vla}{\vec{\lambda}}
\nc{\Nobs}{N_{\mathrm{obs}}}
\nc{\av}[1]{\langle #1 \rangle} 
\nc{\km}{\mathrm{km}}
\nc{\Mpc}{\mathrm{Mpc}}
\nc{\Tobs}{T_{\mathrm{obs}}}
\nc{\Ntemp}{N_{\mathrm{temp}}}
\nc{\fyr}{f_{\mathrm{yr}}}
\nc{\addref}{[\textcolor{red}{add ref}] } 
\nc{\eg}{\textit{e.g.~}}
\nc{\app}{\approx}
\nc{\hf}{\frac{1}{2}}
\nc{\discuss}{\textcolor{red}{Add discussion here!}}
\nc{\red}[1]{\textcolor{red}{#1}}
\nc{\hp}{h_+} 
\nc{\hc}{h_{\times}} 
\nc{\Oh}{\hat{\Omega}}
\nc{\vx}{\vec{x}}
\nc{\mh}{\hat{m}}
\nc{\nh}{\hat{n}}
\nc{\zh}{\hat{z}}
\nc{\ph}{\hat{p}}
\nc{\A}[1]{\mathcal{A}_{#1}}
\nc{\Ogw}[1]{\Omega_{\mathrm{#1}}}
\nc{\bn}[1]{\dt{t}_{\text{#1}}}
\nc{\bC}[1]{{C}_{\text{#1}}}
\nc{\NTOA}{N_{\text{TOA}}}
\nc{\Nmode}{{N_{\text{mode}}}}
\nc{\ARN}{A_{\rm{RN}}}
\nc{\gRN}{\gamma_{\rm{RN}}}
\nc{\bS}{{\Sigma}}
\nc{\br}{{r}}
\nc{\bN}{{R}}
\nc{\Agw}{A_\mathrm{GWB}}
\nc{\UCP}{\mathrm{UCP}}
\nc{\TT}{\mathrm{TT}}
\nc{\ST}{\mathrm{ST}}
\nc{\SL}{\mathrm{SL}}
\nc{\VL}{\mathrm{VL}}
\nc{\lpr}{l^{\prime}}
\nc{\mpr}{m^{\prime}}

\nc{\BFST}{$107 \pm 7$}
\begin{document}
	
\title{Harmonic Analysis on Correlation for Gravitational-Wave Backgrounds of Arbitrary Polarization from Interfering Sources in Generic Dispersion Relation}	

\author{Yan-Chen Bi\orcidlink{0000-0002-9346-8715}}
\affiliation{Institute of Theoretical Physics, Chinese Academy of Sciences, Beijing 100190, China}
\affiliation{School of Physical Sciences, University of Chinese Academy of Sciences, No. 19A Yuquan Road, Beijing 100049, China}

\author{Yu-Mei Wu\orcidlink{0000-0002-9247-5155}}
\email{Corresponding author: wuyumei@yzu.edu.cn} 
\affiliation{Center for Gravitation and Cosmology, College of Physical Science and Technology, Yangzhou University, Yangzhou, 225009, China}
    
\author{Qing-Guo Huang\orcidlink{0000-0003-1584-345X}}
\email{Corresponding author: huangqg@itp.ac.cn}
\affiliation{Institute of Theoretical Physics, Chinese Academy of Sciences, Beijing 100190, China}
\affiliation{School of Physical Sciences, 
    University of Chinese Academy of Sciences, 
    No. 19A Yuquan Road, Beijing 100049, China}
\affiliation{School of Fundamental Physics and Mathematical Sciences, Hangzhou Institute for Advanced Study, UCAS, Hangzhou 310024, China}

\begin{abstract}

The Hellings-Downs (HD) correlation serves as the fundamental benchmark for detecting the gravitational-wave background (GWB) in pulsar timing arrays (PTAs) within General Relativity (GR). However, this canonical signature relies on the idealization of a continuum of sources without interference. In realistic astrophysical scenarios dominated by supermassive black hole binaries (SMBHBs), interference between discrete sources induces intrinsic deviations in the spatial correlation, which may mimic or obscure signatures of modified gravity. 
In this work, we derive the closed-form spatial correlation functions for a GWB with arbitrary polarization and generic GW dispersion relations, in the presence of source interference.
Through a rigorous harmonic analysis, we demonstrate that source interference modifies the correlation shape but strictly preserves the lowest non-vanishing multipole moment characteristic of each polarization, specifically the quadrupole for tensor, dipole for vector, and monopole for scalar modes. The truncation at higher-order multipoles is governed by the interplay between pulsar distances and dispersion effects. 
Furthermore, we quantify the statistical degeneracy between interference-induced variation and modified gravity signatures. We conclude that access to only a single realization of the Universe imposes a fundamental theoretical limit on distinguishing modified gravity from GR using spatial correlations alone.

\end{abstract}

\maketitle


\section{Introduction}
\label{sec:introduction}

Despite its profound success, general relativity (GR) may not be the final word on gravity \cite{Stairs:2003eg,Will:2014kxa}. The unification of gravity with quantum mechanics remains one of the most profound challenges in theoretical physics, often motivating extensions to GR that introduce violations of Lorentz invariance, massive gravitons, or additional scalar and vector degrees of freedom~\cite{Will:1997bb,deRham:2014zqa,Bernardo:2023mxc}. Nowadays the direct detection of a gravitational-wave background (GWB) by Pulsar Timing Arrays (PTAs)~\cite{1990ApJ...361..300F} will open a powerful new window to rigorously test these fundamental questions~\cite{NANOGrav:2023gor,EPTA:2023fyk,Wu:2023hsa,Wu:2023dnp}. In this context, the spatial correlation of the GWB serves as the primary observable for discriminating between GR and alternative theories~\cite{Chen:2021wdo,Chen:2023uiz,Bi:2023ewq,Wu:2023rib,Bernardo:2023zna,Bernardo:2023pwt,Bernardo:2023mxc}. The canonical signature of GR is the Hellings-Downs (HD) correlation~\cite{Hellings:1983fr}, determined by the theory's massless graviton and tensorial nature. However, extensions to GR often predict distinct deviations: modified dispersion relations arising from Lorentz violations or massive gravity, and additional polarization modes permitted by general metric theories~\cite{Chamberlin:2011ev,Gair:2015hra,Ezquiaga:2021ler,Liang:2023ary}. Most general metric theories of gravity permit up to six polarization modes: two tensor ($+$, $\times$), two vector ($x$, $y$), and two scalar ($b$, $l$) modes~\cite{Lee_2008,Chamberlin:2011ev,Boitier:2020xfx,Gair:2015hra}. Validating gravity thus hinges on the precise characterization of these subtle correlation signatures.

However, the search for these fundamental deviations is complicated by the intrinsic complexity of the astrophysical signal. The standard HD derivation idealizes the GWB as an isotropic continuum of unpolarized sources. In reality, if the GWB originates from a population of supermassive black hole binaries (SMBHBs), the signal in the nanohertz band is likely emitted by a finite number of sources emitting at closely spaced frequencies, leading to inevitable interference~\cite{Roebber:2016jzl,Allen:2022dzg,Lamb:2025niq}. This interference introduces intrinsic variations in the spatial correlation pattern, commonly termed ``cosmic variance'', such that the resulting correlations in a single realization of our Universe can mimic the signatures of modified gravity~\cite{Wu:2024xkp,Bernardo:2022xzl}. Crucially, similar astrophysical interference is also present in modified gravity scenarios, further obscuring their correlation signatures.
Consequently, any attempts to constrain fundamental physics parameters, i.e., the graviton mass or the presence of extra polarizations, is theoretically incomplete without accounting for the variance introduced by the source interference.



In this work, we establish a unified theoretical framework to address this complexity. We derive the generalized spatial correlation functions for a GWB with arbitrary polarization and a generic dispersion relation, explicitly incorporating interference effects.
We demonstrate that the coupling between fundamental dispersion and astrophysical interference leads to a novel class of realization-dependent correlation curves. By quantifying the statistical distribution of the Legendre coefficients for different polarization modes in a generic dispersion relation, we assess the theoretical limit of distinguishing modified gravity from GR in the presence of interference. Our results provide the necessary theoretical foundation for future precision tests of gravity using PTA data.

The rest of this paper is organized as follows: In Sec.~\ref{sec:formalism}, we show the expression for redshift of pulsars in generic dispersion relation. In Sec.~\ref{sec:correlations}, we present the general formalism for calculating the correlation. In Sec.~\ref{sec:features}, we show the resulting correlation curves for various scenarios. We discuss the implications and conclude in Sec.~\ref{sec:conclusion}. We adopt natural units by setting the speed of light $c = 1$ and the reduced Planck constant $\hbar = 1$. The Earth is placed at the coordinate origin. Unit vectors are denoted by symbols with hats, such as $\hat{{\Omega}}$ and $\hat{{p}}$.

\section{Redshift response in generic dispersion}
\label{sec:formalism}
We begin by considering a generalized dispersion relation for GWs, $\omega = \omega(k)$, where $k$ is the wave-number. Without loss of generality, the GW waveform from the $r$-th source is modeled as
\m
h_r(t, \vec{x}) = A_r e^{-i[2\pi f_r (t - \eta \hat{{\Om}}_r \cdot \vec{x}) + \Phi_r]},
\label{waveform}
\n
where the parameter $\eta \equiv k / \omega(k)$ encodes the modification to the dispersion relation. Here, $A_r$, $f_r$, $\hat{\Omega}_r$ and $\Phi_r$ denote the amplitude, frequency, propagation direction and initial phase of the $r$-th source, respectively. The redshift $z_a(t)$ induced on the pulse arrival time of pulsar $a$, located in the direction $\hat{{p}}_a$, is given by \cite{Maggiore:2018sht}
\m
z_a(t) = - \frac{1}{2} \hat{p}_a^i \hat{p}_a^j
\int_{t-D_{a}}^{t}
dt^{\prime} \[ \frac{\partial h_{ij}(t^{\prime}, \vec{x})}{\partial t^{\prime}} \]_{\vec{x}=(t - t^{\prime}) \hat{p}_a} ,
\label{redshit}
\n
where $D_a$ is the pulsar distance.
The traceless-transverse metric perturbation $h_{ij}(t, \vec{x})$ is expressed as a superposition of all GW source as
\m
h_{ij}(t, \vec{x}) = \sum_r h_r(t, {\vec{x}}) \epsilon^{\rm P}_{ij}({\hat{\Omega}}_r),
\n
where $\epsilon^{\rm P}_{ij}(\hat{\Omega}_r)$ is the complex polarization tensor for mode ${\rm P} \in \{\mathrm{TT, VL, ST, SL}\}$, defined as \cite{Chamberlin:2011ev,OBeirne:2019lwp}
\m
\epsilon^{\mathrm{TT}}_{ij}({\hat{\Omega}}_r) = \epsilon^+_{ij}({\hat{\Omega}}_r) + i \epsilon^{\times}_{ij}({\hat{\Omega}}_r) , & \ & \epsilon^{\mathrm{ST}}_{ij}({\hat{\Omega}}_r) = \epsilon^{ b}_{ij}({\hat{\Omega}}_r),\\
\epsilon^{\mathrm{VL}}_{ij}({\hat{\Omega}}_r) = \epsilon^x_{ij}({\hat{\Omega}}_r) + i \epsilon^{y}_{ij}({\hat{\Omega}}_r) , & \ & \epsilon^{\mathrm{SL}}_{ij}({\hat{\Omega}}_r) = \epsilon^{ l}_{ij}({\hat{\Omega}}_r).
\n
Explicit definitions of the polarization tensors $\epsilon^{+,\times,x,y,b,l}_{ij}$ are provided in \cite{Chamberlin:2011ev,OBeirne:2019lwp}. Substituting the waveform into \Eq{redshit}, the redshift becomes
\m
z_a(t) = - \frac{1}{2} {\hat{p}}_a^i {\hat{p}}_a^j
\int_{t - D_{a}}^{t}
dt^{\prime} \sum_r A_r (-i 2\pi f_r) \mathrm{e}^{-i(2\pi f_r t^{\prime} + \Phi_r)} \mathrm{e}^{i 2\pi f_r \eta
[{\hat{\Omega}}_r \cdot (t-t^{\prime})\hat{p}_{a}]}
\epsilon^{\rm P}_{ij}({\hat{\Omega}}_r).
\label{redshift1}
\n
The exponential term containing ${\hat{\Omega}}_r$ can be expanded using spherical Bessel functions $j_l(x)$ as
\m
\mathrm{e}^{i 2\pi f_r \eta 
[{\hat{\Omega}}_r \cdot (t-t^{\prime})\hat{p}_{a}]}
= 4 \pi \sum_{lm} i^l j_l(2 \pi f_r \eta (t - t^{\prime})) Y_{lm}^{\star}({\hat{\Omega}}_r) Y_{lm}({\hat{p}_a}).
\n
Employing the variable substitution $y = 2\pi f_r \eta (t - t')$, the integral in \Eq{redshift1} simplifies to
\m
z_a(t) = {2\pi i} \sum_r A_r \mathrm{e}^{-i (2\pi f_r t + \Phi_r)} {\hat{p}}_a^i {\hat{p}}_a^j \epsilon^{\rm P}_{ij}({\hat{\Omega}}_r) \sum_{lm} i^l \int_0^{2\pi f_r D_a \eta} \frac{dy}{\eta} e^{i \frac{y}{\eta}} j_l(y) Y_{lm}^{\star}({\hat{\Omega}}_r) Y_{lm}({\hat{p}_a}) ,
\label{redshift2}
\n
\Eq{redshift2} reveals that the redshift naturally admits an expansion in spherical harmonics $Y_{lm}(\hat{{p}}_a)$, with coefficients
\m
a_{lm} = {i} \sum_{r} A_r \mathrm{e}^{-i (2\pi f_r t + \Phi_r)} B_{lm}^{\rm P}(f_r D_a, {\hat{\Omega}}_r) ,
\n
where
\m
B_{lm}^{\rm P}(f_r D_a, \hat{{\Omega}}_r) =  2\pi i^l \int d{\hat{p}}_a {\hat{p}}_a^i {\hat{p}}_a^j \epsilon^{\rm P}_{ij}({\hat{\Omega}}_r) \int_0^{2\pi f_r D_a \eta} \frac{dy}{\eta} e^{i \frac{y}{\eta}} \sum_{LM} j_L(y) Y_{LM}^{\star}({\hat{\Omega}}_r) Y_{LM}({\hat{p}}_a) Y^{\star}_{lm}({\hat{p}}_a) .
\label{eq:Blm}
\n

\section{Spatial correlations for interfering sources}
\label{sec:correlations}
While the preceding derivation characterizes the response of a single pulsar, PTAs typically monitor an array of pulsars, utilizing cross-correlations to disentangle signals from noise. 
As the number of pulsars increases, multiple pairs share similar angular separations. Consequently, the measurable quantity is the ``pulsar-averaged'' correlation, which aggregates contributions from all pairs separated by a fixed angle. This pulsar-averaged correlation is defined as
\m
\langle \overline{z_a(t) z^{\star}_b(t)} \rangle_{p} = \sum_{lm} \sum_{\lpr \mpr} \overline{a_{lm} a^{\star}_{\lpr\mpr} } \langle Y_{lm}({\hat{p}}_a) Y^{\star}_{\lpr\mpr}({\hat{p}}_b) \rangle_{p},
\n
where the overlines represents the time-average and $\langle \rangle_p$ denotes the pulsar average. The time averaged product of the coefficients is given by
\m
\overline{a_{lm} a^{\star}_{\lpr\mpr} } = \sum_{r,s} A_r A_s \mathrm{e}^{-i(\Phi_r - \Phi_s)} \overline{\mathrm{e}^{-i 2\pi (f_r - f_s) t }} B_{lm}^{\rm P}(f_r D_a, {\hat{\Omega}}_r) B_{\lpr\mpr}^{\rm P \star}(f_s D_b, \hat{{\Omega}}_s). 
\label{time_ave_aa}
\n
The term $\overline{\mathrm{e}^{-2\pi i(f_r-f_s)t}}$ dictates the interference structure. When sources emit GWs with relatively disparate frequencies, cross-terms ($r \neq s$) vanish, leaving only diagonal self-interference. However, while PTAs span a broad frequency range ($10^{-9} \sim 10^{-7}$ Hz), each Fourier bin of width $\Delta f \sim 1$ nHz may contain $\mathcal{O}(10^3)$ sources with nearly identical frequencies~\cite{Roebber:2016jzl}. Consequently, numerous sources emit at frequencies $f_r \approx f_s$, preserving non-diagonal mutual interference terms.


The pulsar average acts primarily on the spherical harmonic product as \cite{Allen:2024bnk}
\m
\langle Y_{lm}({\hat{p}}_a) Y^{\star}_{\lpr\mpr}({\hat{p}}_b) \rangle_{p} = \frac{1}{4\pi} \delta_{l\lpr} \delta_{m\mpr} P_l({\hat{p}}_a \cdot {\hat{p}}_b) ,
\label{pulsar_average}
\n
with $P_l$ are the Legendre polynomials of order $l$. Substituting \Eq{redshift2}, \Eq{time_ave_aa} and \Eq{pulsar_average} into the correlation definition yields
\m
\langle \overline{z_a(t) z^{\star}_b(t)} \rangle_{p} = \frac{1}{4\pi}  \sum_{r,s} A_r A_s \mathrm{e}^{-i(\Phi_r - \Phi_s)} \sum_{lm} B_{lm}^{\rm P}(fD_a, \hat{{\Omega}}_r) B_{lm}^{\rm P \star}(fD_b, \hat{{\Omega}}_s) P_l({\hat{p}}_a \cdot {\hat{p}}_b).
\label{zazb}
\n

To evaluate $B_{lm}^{\rm P}(fD, \hat{{\Omega}})$, we utilize the rotational properties of spherical harmonics. Let $\hat{p}_a = (\sin \theta_a \cos \phi_a, \,\, \sin \theta_a \sin \phi_a, \,\, \\  \cos \theta_a)$.
The general form $B_{lm}^{\rm P}(fD, \hat{{\Omega}})$ can be obtained by rotating $B_{lm}^{\rm P}(fD, \hat{{z}})$, in which the GW source lies along the $\hat{{z}}$-axis. The polarization vectors are defined as $\hat{m} = \cos\phi \hat{x} + \sin \phi \hat{y}$ and $\hat{n} = -\sin\phi \hat{x} + \cos\phi \hat{y}$, where $\phi$ is an arbitrary phase angle that determines the relative angle between polarization vectors and coordinate vectors. The rotation is achieved by the Wigner-D matrix $B^{\rm P}_{lm}(fD, \hat{{\Omega}}) = \sum_{m^{\prime}} D^{l}_{m^{\prime} m}(R) B^{\rm P}_{lm}(fD, \hat{{z}})$ with rotation matrix $R(\alpha, \beta, \gamma)$ chosen to align $\hat{z}$ with $\hat{\Omega}$ while preserving polarization orientation~\cite{Allen:2024bnk}. 
Under this configuration, the contractions between the pulsar vector and polarization tensor simplify significantly
\m
{\hat{p}}^i {\hat{p}}^j \epsilon^{\mathrm{TT}}_{ij}({\hat{z}}) = 4 \sqrt{\frac{2\pi}{15}} Y_{22}(\hat{p}) \mathrm{e}^{-i2\phi} , \ {\hat{p}}^i {\hat{p}}^j \epsilon^{\mathrm{ST}}_{ij}({\hat{z}}) = \frac{4\sqrt{\pi}}{3} Y_{00}(\hat{p}) - \frac{4}{3} \sqrt{\frac{\pi}{5}} Y_{20}(\hat{p}) ,
\label{contra1}\\ 
{\hat{p}}^i {\hat{p}}^j \epsilon^{\mathrm{VL}}_{ij}({\hat{z}}) = - 2 \sqrt{\frac{2\pi}{15}} Y_{21}(\hat{p}) \mathrm{e}^{-i\phi} , \
{\hat{p}}^i {\hat{p}}^j \epsilon^{\mathrm{SL}}_{ij}({\hat{z}}) = \frac{2\sqrt{\pi}}{3} Y_{00}(\hat{p}) + \frac{4}{3} \sqrt{\frac{\pi}{5}} Y_{20}(\hat{p}) .
\label{contra2}
\n
Consequently, the $B_{lm}^{\rm P}(fD_a, \hat{{z}})$ are derived as
\m
B_{lm}^{\rm TT}(fD, \hat{{z}}) &=& - \delta_{m2} 2\sqrt{2} \pi  i^l \mathrm{e}^{-i 2\phi} \sqrt{\frac{(l+2)!}{(l-2)!} \frac{2l+1}{8\pi}} \int_0^{2\pi f D \eta } \frac{dy}{\eta} \mathrm{e}^{i\frac{y}{\eta}} \frac{j_l(y)}{y^2} , \label{blmttz} \\
B_{lm}^{\rm ST}(fD, \hat{{z}}) &=& \delta_{m0} 2 \pi  i^l \mathrm{e}^{-i \phi} \sqrt{\frac{2l+1}{4\pi}} \int_0^{2\pi f D \eta} \frac{dy}{\eta} \mathrm{e}^{i\frac{y}{\eta}} \(\frac{d^2 j_l(y)}{dy^2} + j_l(y)\) , \label{blmstz}\\
B_{lm}^{\rm VL}(fD, \hat{{z}}) &=& -\delta_{m1} 4 \pi  i^l \mathrm{e}^{-i \phi} \sqrt{l(l+1)} \sqrt{\frac{2l+1}{4\pi}} \int_0^{2\pi f D \eta} \frac{dy}{\eta} \mathrm{e}^{i\frac{y}{\eta}} \frac{d}{dy} \frac{j_l(y)}{y} , \label{blmvz} \\
B_{lm}^{\rm SL}(fD, \hat{{z}}) &=& - \delta_{m0} 2 \pi  i^l \mathrm{e}^{-i \phi} \sqrt{\frac{2l+1}{4\pi}} \int_0^{2\pi f D \eta} \frac{dy}{\eta} \mathrm{e}^{i\frac{y}{\eta}} \frac{d^2 j_l(y)}{dy^2}  \label{blmslz}.
\n
The general forms $B_{lm}^{\rm P}$ for arbitrary GW directions $\hat{\Omega}$ are given then
\m
B_{lm}^{\rm TT}(fD, \hat{{\Omega}}) &=& - 2\pi i^l _{-2}Y^{\star}_{lm}(\hat{{\Omega}}) \sqrt{\frac{(l+2)!}{(l-2)!}}  I_l^{\rm TT}(fD,\eta) , \quad 
B_{lm}^{\rm ST}(fD, \hat{{\Omega}}) = 2 \pi i^l Y^{\star}_{lm}(\hat{{\Omega}}) I_l^{\rm ST}(fD,\eta) , \\
B_{lm}^{\rm VL}(fD, \hat{{\Omega}}) &=& - 4 \pi i^l _{-1}Y^{\star}_{lm}(\hat{{\Omega}}) \sqrt{l(l+1)} I_l^{\rm VL}(fD,\eta) , \quad
B_{lm}^{\rm SL}(fD, \hat{{\Omega}}) = - 2 \pi i^l Y^{\star}_{lm}(\hat{{\Omega}}) I_l^{\rm SL}(fD,\eta) ,
\n
where $I_l^{\rm P}(fD,\eta)$ denotes the integrals over $y$ appearing in Eqs.~\Eqs{blmttz}-\Eqs{blmslz}.

\section{Correlation Features}
\label{sec:features}
Substituting the explicit expressions for $B_{lm}^{\rm P}(fD, \hat{{\Omega}})$ into Eq.~\eqref{zazb}, we derive the closed-form correlation function, which admits a Legendre polynomial expansion as $\langle \overline{z_a(t) z^{\star}_b(t)} \rangle_{p} = \sum_l w^{\rm P}_l P_l({\hat{p}}_a \cdot {\hat{p}}_b)$,
where the coefficients $w_l^{\rm P}$ are defined as
\m
w_l^{\rm P} = q_l^{\rm P} I_l^{\rm P}(fD_a,\eta) I_l^{\rm P \star}(fD_b,\eta) \sum_{r,s} \mathcal{A}_r\mathcal{A}_{s} \mathrm{e}^{-i(\phi_r - \phi_s)} \mathcal{S}^P_l(\hat{\Omega}_r, \hat{{\Omega}}_s). 
\label{w_l}
\n
Here, we employ the normalized amplitudes $\mathcal{A}_r\mathcal{A}_{s} = A_r A_s / \sum_n A_n^2$ to remove spectral dependence. 
Detailed expressions for the factors $q^{\rm P}_l$ and the functions $\mathcal{S}^P_l(\hat{\Omega}_r, \hat{{\Omega}}_s)$ are provided in Table~\ref{tab:coefficient},
where $\cos\beta_{rs} = \hat{{\Omega}}_r \cdot \hat{{\Omega}}_s$ denotes the angular separation between two GW sources,  $J_{\ell}^{(\mu,\nu)}(\cos\beta)$ is the Jacobi Polynomials of degree $\ell$, and the phase factor $\chi(\hat{\Omega}_r, \hat{\Omega}_s)$ is defined by the relation~\cite{castillo20033,doi:10.1142/0270}
\m
\tan \frac{1}{2} \chi(\hat{\Omega}_r, \hat{\Omega}_s) = \frac{\sin\frac{1}{2}(\phi_r - \phi_s) \cos\frac{1}{2}(\theta_s + \theta_r)}{\cos\frac{1}{2}(\phi_s - \phi_r) \cos\frac{1}{2}(\theta_s - \theta_r)} ,
\n
with $\theta_r$ and $\phi_r$ denoting the polar and azimuth angles of the propagation direction $\hat{\Omega}_r$ for $r$-th source, respectively.  \Eq{w_l} evidently shows that the Legendre coefficients $w_l^{\rm P}$ can factorize into a deterministic GW-propagation-dependent component (governed by $I_l^{\rm P}$) and a stochastic source-dependent component (the sum over $r,s$). The former depends only on inherently-known pulsar distances and the dispersion relation, whereas the latter originating from interference effect depends on the inherently-unknown physical realization of the SMBHB population, in particular the initial phases and sky locations of individual SMBHBs.

\begin{table*}[h]
    \centering
    \setlength{\tabcolsep}{14pt}
    \caption{The Legendre coefficients for arbitrary modes. \label{tab:coefficient}}
    \label{prior}
    \begin{tabular}{ccc}
        \hline\hline
        \textbf{GW modes P} & $q^{\rm P}_l$ & $\mathcal{S}^P_l(\hat{\Omega}_r, \hat{{\Omega}}_s)$  \\
        \hline
        {TT} & ${(2l+1)(l+2)!}/{(4(l-2)!)}$ & $\cos^4 ({\beta_{rs}}/{2}) J^{(0,4)}_{l-2}(\cos\beta_{rs}) \mathrm{e}^{2i\chi(\hat{\Omega}_r, \hat{\Omega}_s)}$  \\
        \hline
        {VL} & $l(l+1)(2l+1)$ & $\cos^2 ({\beta_{rs}}/{2}) J^{(0,2)}_{l-1}(\cos\beta_{rs}) \mathrm{e}^{i\chi(\hat{\Omega}_r, \hat{\Omega}_s)}$  \\
        \hline
        {ST} & ${(2l+1)}/{4}$ & $P_l(\cos\beta_{rs})$  \\
        \hline
        {SL} & ${(2l+1)}/{4}$ & $P_l(\cos\beta_{rs})$  \\
        \hline\hline
    \end{tabular}
\end{table*}

Although the Legendre coefficients $w_l^P$ are quantitatively affected by the interference terms determined by the detailed correlation pattern, they preserve the lowest non-vanishing multipole order characteristic of each polarization mode. This preservation, namely quadrupole ($l=2$) for tensor modes, dipole ($l=1$) for vector modes, and monopole ($l=0$) for scalar modes, is ensured by the contraction of the polarization patterns in \Eq{eq:Blm}, \Eq{contra1} and \Eq{contra2} and remains intact even when interference effects are fully incorporated.

In the infinite pulsar distance limit ($fD \to \infty$) which is commonly adopted in PTA analyses as a “long-arm” approximation, the integrals $I_l^{\rm P}(fD, \eta)$ simplify to the following analytical forms \cite{Qin:2020hfy}
\m
I_l^{\rm TT}(\infty,\eta) &=& \frac{1}{\eta} I_l^{(2)}(\eta), \qquad \qquad \qquad \qquad
I_l^{\rm ST}(\infty,\eta) = \frac{i}{\eta^2} \delta_{l0} - \frac{1}{3\eta} \delta_{l1} + \frac{(\eta^2 - 1)}{\eta^3}I_l^{(0)}(\eta),\\
I_l^{\rm VL}(\infty,\eta) &=& -\sqrt{\frac{2}{l(l+1)}} \frac{1}{3\eta} - \frac{i}{\eta^2} I_l^{(1)}(\eta), \ 
I_l^{\rm SL}(\infty,\eta) = \frac{i}{\eta^2} \delta_{l0} - \frac{1}{3\eta} \delta_{l1} - \frac{1}{\eta^3} I_l^{(0)}(\eta),
\n
with the auxiliary functions defined as
\m
I_l^{(n)}(\eta) = \left\{ 
\begin{aligned}
& \(\frac{i}{2\eta}\)^{l+1-n} \frac{\sqrt{\pi}}{2^n} \frac{\Gamma(l+1-n)}{\Gamma(l+\frac{3}{2})} \ _2F_1(\frac{l+1-n}{2}, \frac{l+2-n}{2}, l+\frac{3}{2}, \eta^{-2}), \ \mathrm{for} \ \eta < 1, \\ 
& i^{l+1-n} 2^{n-1} (n-1)! \frac{(l-n)!}{(l+n)!} , \ \mathrm{for} \ \eta = 1, \\
& 2^{-(n+1)} \sqrt{\pi} \Bigg{\{} \frac{\Gamma(\frac{l+1-n}{2})}{\Gamma(\frac{l+2+n}{2})} \ _2F_1(-\frac{l+n}{2}, \frac{l+1-n}{2}, \frac{1}{2}, \eta^{2}) - \frac{2i}{\eta(2l+1)} \frac{\Gamma(\frac{l+2-n}{2})}{\Gamma(\frac{l+1+n}{2})}  \\
& \times \[\ _2F_1(-\frac{l+1+n}{2}, \frac{l+2-n}{2}, \frac{1}{2}, \eta^{2}) - \ _2F_1(-\frac{l-1+n}{2}, \frac{l-n}{2}, \frac{1}{2}, \eta^{2}) \] \Bigg{\}} , \ \mathrm{for} \ \eta > 1.
\end{aligned}
\right.
\n
As expected, in the absence of interference effects, the above expressions recover the standard correlation curves after setting $\eta = 1$ and $fD \to \infty$. 

\begin{figure}[htpb]
\centering
\includegraphics[width=1.\textwidth]{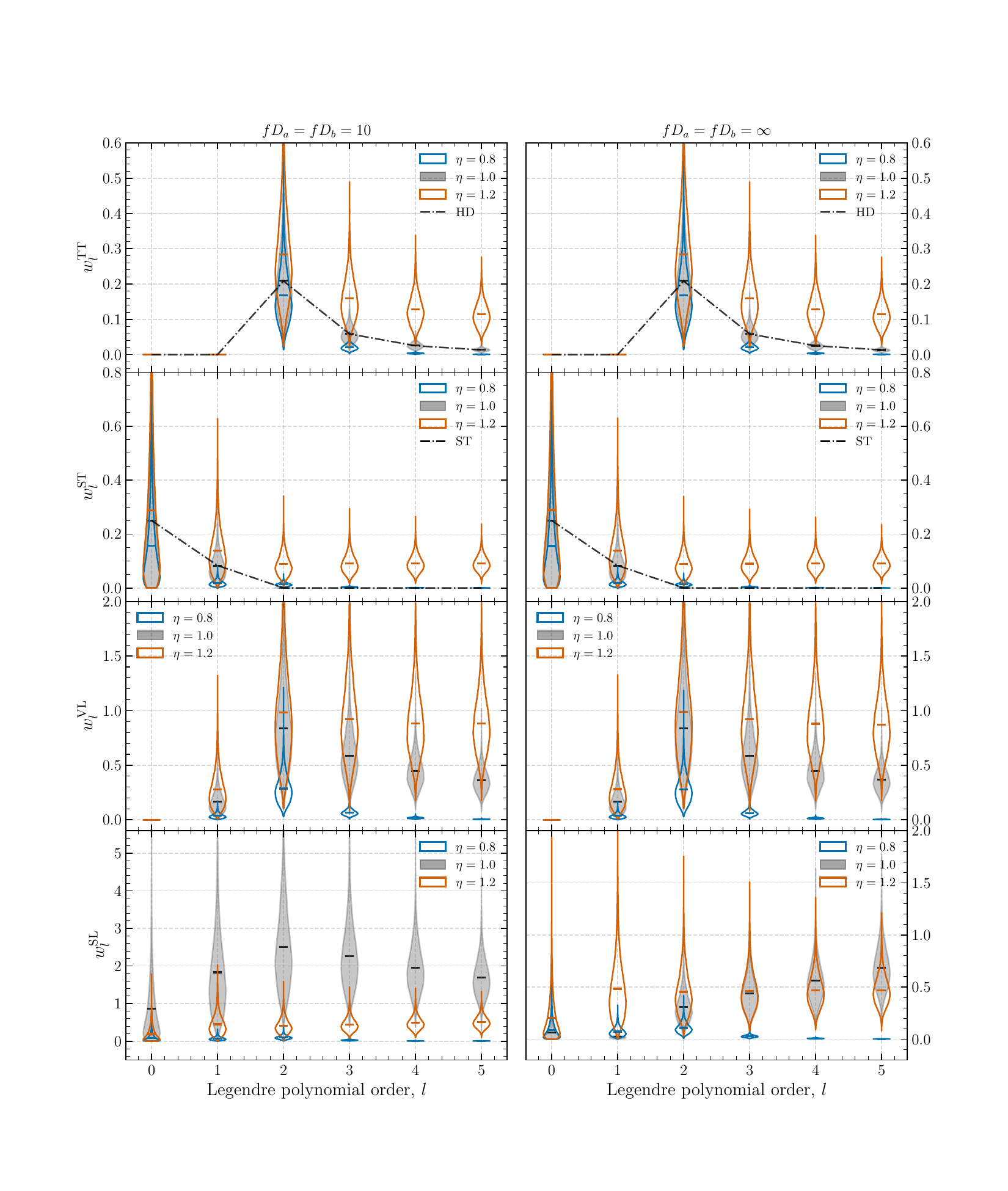}
\caption{Distributions of the normalized Legendre coefficients $w_l^{\rm P}$ for all polarization modes, derived from 10,000 stochastic realizations across distinct dispersion parameters $\eta$ and pulsar configurations. The left and right columns illustrate results for the finite-distance ($fD_a = fD_b = 10$) and infinite-distance ($fD_a = fD_b = \infty$) cases, respectively. Blue, grey, and orange violins denote $\eta = 0.8$, $1.0$, and $1.2$, respectively. For comparison, the dash-dotted lines in the first and second rows represent the analytical predictions for the TT and ST modes. \label{wl_distribution}}
\end{figure}

In terms of the interference effect, since the specific realization of the SMBHB population is unknown, we employ numerical simulations to characterize the statistical properties of the correlation~\cite{Wu:2024xkp,Wu:2025xqh}. 
we generate an ensemble of 10,000 stochastic realizations, each comprising 1,000 unpolarized monochromatic GW sources. The source waveforms follow the prescription in Eq.~\eqref{waveform}, with amplitudes modeled as $A_j = j^{-1/3}$ ($j$ being the source index) to represent a spatially uniform source population.\footnote{Note that the normalized Legendre coefficients are independent of the absolute scale of $A_j$~\cite{Wu:2024xkp,Wu:2025xqh}.} Propagation directions $\hat{\Omega}_j$ are distributed isotropically across the celestial sphere, while initial phases $\phi_j$ are drawn uniformly from the interval $[0, 2\pi)$. Variations in source configurations across realizations produce different correlation patterns. Figure~\ref{wl_distribution} presents the resulting distributions of the normalized Legendre coefficients for all polarization modes. The harmonic analysis is truncated at multipole moment $l=5$, commensurate with the projected angular resolution of near-future PTA experiments~\cite{NANOGrav:2023tcn}. For reference, the dash-dotted lines in the first and second rows respectively standard predictions for the TT and ST modes ($\eta = 1$, $fD \to \infty$, and in the absence of interference effects), for which closed-form expressions exist and are given by
\begin{equation}
w^{\rm TT, analytical}_l = \left\{
\begin{aligned}
    &0 , &\text{for} & \ l=0,1 \\
    &(2l+1) \frac{(l-2)!}{(l+2)!} , &\text{for} & \ l \geq 2
\end{aligned}
\right. \text{and,} \quad
w^{\rm ST, analytical}_l = \left\{
\begin{aligned}
    &\frac{1}{4} , &\text{for} & \ l=0 \\
    &\frac{1}{12} , &\text{for} & \ l=1 \\
    &0 , &\text{for} & \ l \geq 2
\end{aligned}
\right. .
\end{equation}

A comparison between finite-distance scenarios (left column) and the infinite-distance limit (right column) reveals excellent agreement for most modes, with the notable exception of the SL mode. This discrepancy underscores the necessity of incorporating finite-distance effects when analyzing the SL mode. 
It is worth noting that finite pulsar distances also introduce corrections for TT, ST, and VL modes at small pulsar separations, i.e. corresponding to higher multipoles $l$. However, within the limited multipole range considered here, these effects are negligible.


While previous investigations~\cite{Bernardo:2023pwt,Bernardo:2022rif,Qin:2020hfy} were largely restricted to the parameter space defined by $\eta < 1$, the present study provides an extended analysis across a more comprehensive range of dispersion parameters. In the following, we concentrate on the physical regime in the immediate vicinity of the GR limit ($\eta = 1$), whereas a broader exploration of the parameter space is relegated to Appendix~\ref{sec:appa}.

In all panels of Figure~\ref{wl_distribution}, grey violins represent the luminal case ($\eta=1$), while blue and orange violins denote $\eta=0.8$ and $\eta=1.2$, respectively. Modified dispersion relations ($\eta \neq 1$) significantly modulate the spectral distribution of the Legendre coefficients $w_l^{\rm P}$. Specifically, for the TT, ST and VL modes, sub-luminal propagation ($\eta < 1$) systematically suppresses the ensemble variance (compressing the statistical spread), whereas super-luminal propagation ($\eta > 1$) broadens it. 
In contrast, the SL mode manifests a more intricate dependence on the dispersion parameter $\eta$ and the pulsar distance. In the ``long-arm" limit, variations in the statistical spread relative to the $\eta=1$ case show a strong dependence on the Legendre multipole order $l$ for both sub-luminal and super-luminal propagation.
By contrast, in the finite-distance case, both sub-luminal and super-luminal dispersion lead to a pronounced attenuation of the ensemble variance.


Interference between GW sources introduces realization-dependent variations in correlation patterns, complicating the distinction between modified gravity and GR. While a modified dispersion relation ($\eta \neq 1$) may reduce or extend the ensemble spread of Legendre coefficients, we remain limited to a single realization of the Universe, precluding access to the full ensemble distribution~\cite{Wu:2024xkp,Wu:2025xqh,Romano:2023zhb,Allen:2024uqs,Pitrou:2024scp}. Consequently, for any particular realization, the overlapping distributions prevent definitive discrimination between modified gravity and interference effects within GR.

Among the various polarization channels, transverse modes merit particular attention as they exhibit favorable Bayes factors in contemporary analyses~\cite{Chen:2023uiz}. It is noteworthy that while the scalar transverse (ST) mode vanishes for $l \geq 2$ in the standard luminal case ($\eta=1$)~\cite{Qin:2020hfy,Bernardo:2022rif}, a generalized dispersion relation ($\eta \neq 1$) excites non-vanishing ST power at higher multipoles. This phenomenon hinders the discrimination between distinct transverse polarizations in harmonic space. Nevertheless, under the constraint of luminal propagation, these modes remain strictly separable.

\begin{figure}[htpb]
\centering
\includegraphics[width=1.\textwidth]{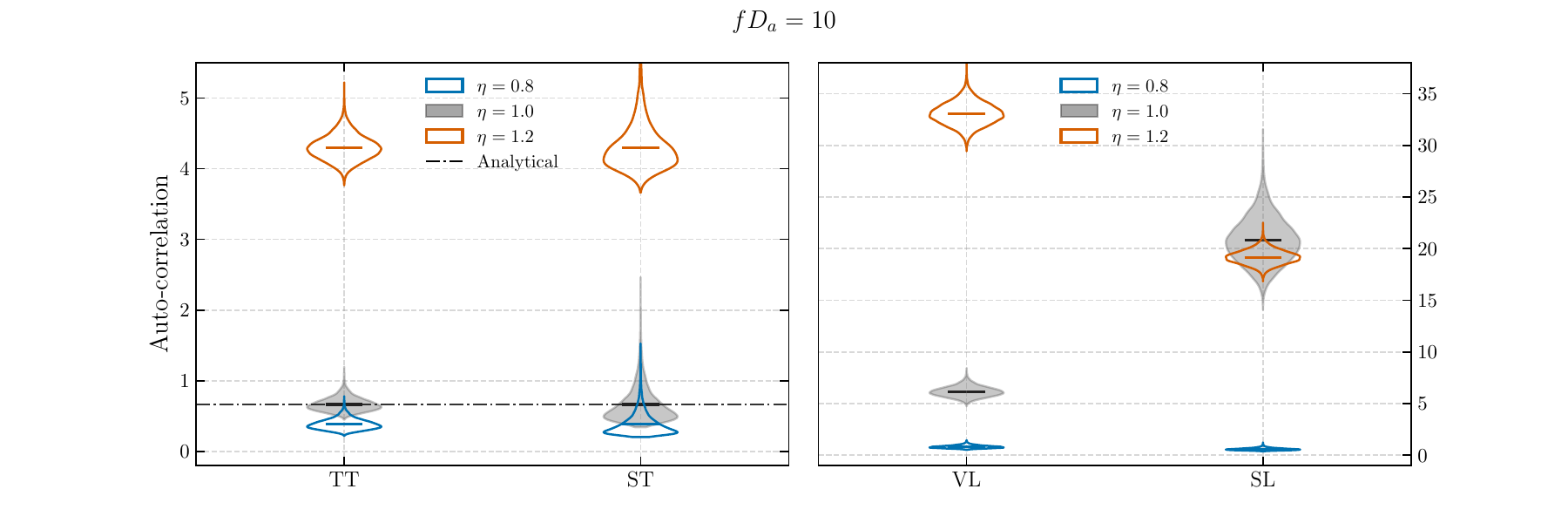}
\caption{Distributions of pulsar auto-correlations for all polarization modes, derived from 10,000 stochastic realizations across distinct dispersion parameters $\eta \in \{0.8, 1.0, 1.2\}$. All distributions are computed for a fixed pulsar configuration of $fD_a = 10$. Blue, grey, and orange violins denote the cases for $\eta = 0.8$, $1.0$, and $1.2$, respectively. For reference, the black dash-dotted line denotes the analytical predictions for the TT and ST modes. \label{auto-correlation}}
\end{figure}


We complete our discussion by calculating the auto-correlations, $\Gamma_{aa}$, which play a particularly important role in PTA analyses as they characterize the total power induced by the GWB in an individual pulsar's response. These terms are evaluated as
\m
\Gamma_{aa} = \langle \overline{z_a(t) z^{\star}_a(t)} \rangle_{p} = \sum_l q_l^{\rm P} |I_l^{\rm P}(fD_a,\eta)|^2 \(1 + \sum_{r \neq s} \mathcal{A}_r\mathcal{A}_{s} \mathrm{e}^{-i(\phi_r - \phi_s)} \mathcal{S}^P_l(\hat{\Omega}_r, \hat{{\Omega}}_s) \).
\n
As in the case of cross-correlations, the resulting mathematical structure of the auto-correlation reflects the superposition of the original contributions and additional interference terms, consistent with the framework established in Ref.~\cite{Wu:2024xkp}. The distributions of pulsar auto-correlations for all polarization modes obtained from stochastic realizations are illustrated in Fig.~\ref{auto-correlation}. Blue, grey, and orange violins denote the cases $\eta = 0.8$, $1.0$, and $1.2$, respectively.
The black dash-dotted line represents the analytical predictions for the TT and ST modes, given by $\Gamma_{aa} = 2\sum_l w_l^{\rm TT/ST,,analytical} = 2/3$\footnote{Most existing works adopt $\Gamma_{aa}=1$, which merely reflects a different normalization convention.}. Notably, since the auto-correlations for the VL and SL modes diverge in the limit of infinite pulsar distance~\cite{Hu:2022ujx}, the present analysis focuses exclusively on the physically relevant finite-distance regime. As can be seen, the dispersion parameter $\eta$ modulates the interference-induced spread of the auto-correlation in a manner analogous to its effect on the Legendre coefficients of the cross-correlations.

\section{Conclusion}
\label{sec:conclusion}

In the nanohertz frequency regime, the superposition of multiple GW sources inevitably creates an interfering background, causing the observed spatial correlation in PTAs to deviate from previous predictions. We have presented a systematic harmonic analysis of GWB correlations with arbitrary polarization and a generic dispersion relation, explicitly incorporating source interference.
Treating each polarization mode independently, we derive analytical closed-form expressions for the Legendre coefficients of the spatial correlation. These expressions naturally separate into two structurally distinct components: GW-propagation-related terms and source-related contributions. 
A key finding is that interference preserves the lowest non-vanishing multipole moment for each polarization, while the truncations at higher multipoles are governed by pulsar distances and dispersion parameters.

Through extensive simulations of different SMBHB realizations, we obtain the statistical distributions of Legendre coefficients $w_l^P$. Our simulations demonstrate that source interference introduces realization-dependent variations in the correlation pattern, contingent on the specific configuration of SMBHB positions and phases. This variability complicates the discrimination between modified gravity and general relativity. In addition, both the pulsar distance and a modified dispersion relation ($\eta \neq 1$) systematically modulate the ensemble variance of the Legendre coefficients, leading to either a suppression or an enhancement of the statistical spread. This modulation of the correlation power has important implications for PTA data analysis.

Specifically, the substantial overlap between the distributions for $\eta=0.8, 1.0$ and $1.2$ reveals a fundamental degeneracy: for a single realization of the Universe, a deviation in the observed (auto-)correlation power could be interpreted either as a signature of modified gravity or merely as a statistical fluctuation arising from source interference within GR, even in the idealized limit of noise-free pulsars. This result underscores the necessity of moving beyond the ensemble average expectations and incorporating realization-specific likelihoods in future gravity tests.



Finally, we emphasize that these interference effects predominantly apply to astrophysical backgrounds. For a cosmological GWB \cite{Chen:2019xse,Chen:2022azo,Wu:2023hsa}, the ergodic hypothesis provides access to ensemble statistics through spatial averaging, eliminating the variance and thus approaching the standard averaged correlation~\cite{Caprini:2018mtu}.


\begin{acknowledgments}

YMW is supported by the National Natural Science Foundation of China (Grant No.~12505086). QGH is supported by the National Natural Science Foundation of China (Grant No.~12547110,~12475065,~ 12447101) and the China Manned Space Program with grant no. CMS-CSST-2025-A01.
	
\end{acknowledgments}

\appendix
\section{Broader exploration of the parameter space $\eta$}
\label{sec:appa}

\begin{figure}[htpb]
\centering
\includegraphics[width=0.8\textwidth]{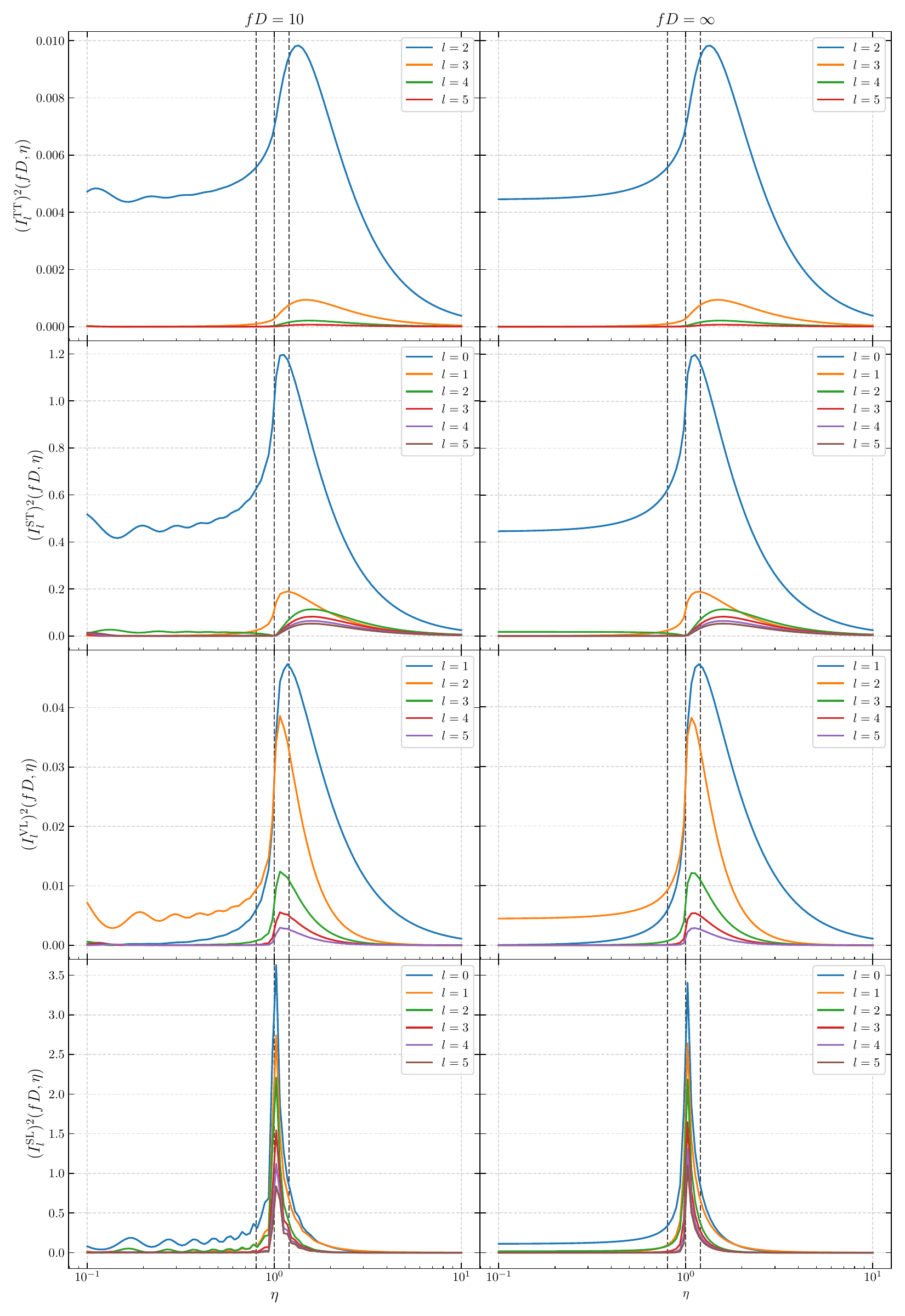}
\caption{Comparison of the function $(I_l^P)^2(fD, \eta)$ between the finite-distance regime $fD=10$ and the infinite-distance limit. The vertical dashed lines denote the dispersion parameters $\eta = 0.8, 1.0$, and $1.2$ from left to right. \label{Il}}
\end{figure}

In this Appendix, we present an extended investigation of the function $(I^{P}_l)^2(fD, \eta)$, defined as the square of $I^{P}_l(fD, \eta)$, over a broader range of the dispersion parameter $\eta$. In the sub-luminal regime ($\eta < 1$), the function $(I^{P}_l)^2(fD, \eta)$ increases monotonically as $\eta$ approaches the luminal limit. In contrast, the super-luminal regime ($\eta > 1$) exhibits more complex behavior. As $\eta$ increases, the integral initially grows before undergoing a rapid decline. The functional dependence of $(I^{P}_l)^2(fD, \eta)$ is illustrated in \Fig{Il}, where the left and right columns correspond to the finite-distance regime ($fD=10$) and the infinite-distance limit, respectively. The vertical dashed lines mark the benchmark dispersion parameters $\eta = 0.8, 1.0$, and $1.2$. Notably, the oscillatory features present in the sub-luminal regime are suppressed in the infinite-distance limit.

\bibliography{./ref}

\end{document}